\documentclass[aip,apl,reprint]{revtex4-1}
\usepackage{epsfig,amsmath,amssymb}
\newcommand {\rhovec}{\ensuremath \boldsymbol{\rho}}
\begin{document}
\title{Comment on ``Turbulence-free ghost imaging'' [Appl. Phys. Lett. 98, 111115 (2011)]} 
\author{Jeffrey H.\ Shapiro}
\affiliation{$^1$Research Laboratory of Electronics, Massachusetts Institute of Technology, Cambridge, MA 02139 USA}
\date{\today}

\maketitle

Meyers, Deacon, and Shih\cite{APL} claim that lensless pseudothermal ghost imaging is immune to the presence of atmospheric turbulence, i.e., it does not suffer the loss of spatial resolution that turbulence imposes on imagery formed with conventional cameras.  In support of that claim they present results from laboratory experiments and theory based on nonlocal two-photon interference. Prior theory\cite{Cheng} for lensless pseudothermal ghost imaging of transmissive objects in the presence of atmospheric turbulence predicts that there \em will\/\rm\ be spatial resolution loss when the source diameter exceeds the source-plane coherence length of the turbulence.  Although the experiments of Meyers, Deacon, and Shih were performed in reflection, using a rough-surfaced (quasi-Lambertian) object, Hardy and Shapiro\cite{Hardy} have shown, theoretically, that such an arrangement also suffers spatial resolution loss when the source diameter exceeds the source-plane coherence length of the turbulence.  References~2 and 3 employed semiclassical photodetection theory, instead of the quantum theory of Meyers, Deacon, and Shih, in their analyses of lensless pseudothermal ghost imaging.  Shapiro and Boyd\cite{SBqip}, however, have unified the semiclassical treatment of ghost imaging in the absence of turbulence with the corresponding nonlocal two-photon interference approach preferred by Meyers, Deacon, and Shih.  In what follows we will refute the turbulence immunity claimed by Meyers, Deacon, and Shih by showing:  (1) their experiments were performed in the regime wherein the source diameter was considerably smaller than the source-plane coherence length of the turbulence; and (2) their theory is only valid in that same regime.  In short, nothing in Ref.~1 contradicts the predictions from Refs.~2 and 3 about the onset of turbulence-induced spatial resolution loss in lensless pseudothermal ghost imaging.  

The source-plane turbulence coherence length, $\rho_0$, is given by \cite{Shapiro}
\begin{equation}
\rho_0 = \left(2.91 k^2\int_0^L\!dz\, C_n^2(z)(1-z/L)^{5/3}\right)^{-3/5},
\label{rho0}
\end{equation} 
where, $k=2\pi/\lambda$ is the wave number at the laser wavelength $\lambda$, and $C_n^2(z)$ is the turbulence strength profile along the propagation path from the source plane $(z=0)$ to the object plane $(z=L)$.  Meyers, Deacon, and Shih do \em not\/\rm\ report their laser wavelength, \em nor\/\rm\ do they report---from inference or direct measurement---their source-plane turbulence coherence length.  They do provide the source diameter (11\,mm), source-to-object path length (1.4\,m), and the maximum turbulence strength ($C_n^2 = 1.5\times 10^{-12}\,{\rm m}^{-2/3}$) although they do not describe its distribution along the propagation path.  We shall assume $\lambda = 780$\,nm, as the Shih group employed in Ref.~6.  We shall also assume that the turbulence is uniformly distributed ($C_n^2 = $ constant) along the $z=0$ to $z=L$ propagation path from the source to the object, because this is the worst (minimum $\rho_0$) case.  We then find that $\rho_0$ = 5\,cm, which places the worst-case experiments from Ref.~1 squarely in the regime wherein Ref.~3 states there will \em not\/\rm\ be any loss of spatial resolution in the lensless pseudothermal ghost image.  

To support their experimental observations of turbulence immunity, Meyers, Deacon, and Shih sketch a theoretical justification for this behavior.  We can identify the failing of their turbulence-immunity theory from their Eq.~(3).  Here they employ phase fluctuations $\Delta\varphi_1(\vec{\rho}_1)$ and $\Delta \varphi_2(\vec{\rho}_2)$ to represent the turbulence-induced propagation effects on the source-to-object-to-bucket-detection at spatial position $\vec{\rho}_1$ and the source-to-speckle-plane-to-speckle-plane-detection at spatial position $\vec{\rho}_2$, respectively.  Aside from this approach's suppressing turbulence-induced scintillation, its major failing is that it has no explicit dependence on the source-plane coordinates.  In particular, the left-hand side of their Eq.~(3) should be
\begin{align}
&|g_2(\vec{\rho}_2,z_2,\vec{\kappa}\,)e^{i\Delta\varphi_2(\vec{\rho}_2,\vec{\kappa}\,)}
g_1(\vec{\rho}_1,z_1,\vec{\kappa}')e^{i\Delta\varphi_1(\vec{\rho}_1,\vec{\kappa}')} \nonumber \\
&+ g_2(\vec{\rho}_2,z_2,\vec{\kappa}')e^{i\Delta\varphi_2(\vec{\rho}_2,\vec{\kappa}')}
g_1(\vec{\rho}_1,z_1,\vec{\kappa}\,)e^{i\Delta\varphi_1(\vec{\rho}_1,\vec{\kappa}\,)}|^2, \nonumber
\end{align}
where the additional (source-plane $\vec{\kappa}$ and $\vec{\kappa}'$) arguments in the turbulence-induced phase fluctuations account for the source-plane turbulence dependence in the extended Huygens-Fresnel principle \cite{Cheng,Hardy}.  Now, it is apparent that the turbulence cancellation claimed by Meyers, Deacon, and Shih in their Eq~(3) \em only\/\rm\ occurs when we can say that 
$\Delta\varphi_2(\vec{\rho}_2,\vec{\kappa}\,) \approx \Delta\varphi_2(\vec{\rho}_2,\vec{\kappa}')$,
and
$\Delta\varphi_1(\vec{\rho}_1,\vec{\kappa}\,) \approx \Delta\varphi_1(\vec{\rho}_1,\vec{\kappa}')$
over the relevant ranges of the arguments.  To show that the validity of these conditions is equivalent to saying the source diameter is smaller than the source-plane turbulence coherence length, we shall extend the nonlocal two-photon interference theory from Ref.~4 to include the presence of turbulence.\cite{footnote} 

From Eqs.~(69) and (71) of Shapiro and Boyd we have that
\begin{align}
&(\pi\rho_0^2)^4|E_m(t-L/c)h_b(\rhovec_b,\rhovec_m)E_{m'}(t-L/c)h_p(\rhovec_p,\rhovec_{m'}) \nonumber\\
&+E_{m'}(t-L/c)h_b(\rhovec_b,\rhovec_{m'})E_m(t-L/c)h_p(\rhovec_p,\rhovec_m)|^2 \nonumber
\end{align}
plays the role of the left-hand side of Eq.~(3) from Ref.~1.  Here:  $E_k(t)$, for $k=m,m'$, is the temporal behavior of the $k$th subsource in the $z=0$ plane; $\rhovec_b$ is a point on the bucket detector; $\rhovec_p$ is a point on the reference detector; and
\begin{equation}
h_j(\rhovec',\rhovec) \equiv \frac{e^{ikL + ik|\rhovec'-\rhovec|^2/2L}}{i\lambda L}
e^{\psi_j(\rhovec',\rhovec)},
\end{equation}
gives the extended Huygens-Fresnel principle Green's function for the source-to-bucket ($j$=$b$) and source-to-reference ($j$=$p$) paths, with $\psi_j(\rhovec',\rhovec)$ being the complex-valued (log-amplitude and phase) fluctuation imposed by the turbulence on path $j$.    The relevant second-order Glauber coherence function for determining the spatial resolution behavior of the lensless pseudothermal ghost image is the ensemble average of the expression given above Eq.~(2).  Exploiting the statistical independence of the different subsources, their statistical independence from the turbulence fluctuations, and assuming that the turbulence on the two propagation paths are statistically independent and identically distributed, we find that the second-order Glauber coherence function is 
\begin{align}
&2\left(\frac{\pi\rho_0^2}{\lambda L}\right)^4\!\langle |E_m(t-L/c)|^2\rangle|\langle |E_{m'}(t-L/c)|^2\rangle \nonumber \\ 
&\times \left[1 + {\rm Re}\!\left(e^{ik(\rhovec_b-\rhovec_p)\cdot(\rhovec_m-\rhovec_{m'})/L}e^{-|\rhovec_m-\rhovec_{m'}|^2/\rho_0^2}\right)\right], \nonumber
\end{align}
where we have used the square-law approximation in evaluating the mutual coherence functions of the turbulence terms, cf.\ Eqs.~(77) and (79) of Ref.~4.  As noted by Meyers, Deacon, and Shih, the constant-in-space term in the second-order Glauber coherence function is responsible for the featureless background component seen in lensless pseudothermal operation, whereas the other (spatially-varying) term gives rise to the ghost image.  Our nonlocal two-photon interference evaluation of that coherence function clearly shows that turbulence immunity will \em only\/\rm\ exist when $\max_{m,m'}|\rhovec_m-\rhovec_{m'}| < \rho_0$.  But the left-hand side of this inequality equals the source diameter, so we have shown that the turbulence-immunity proof from Ref.~1 is limited to that regime.

This work was supported by the DARPA Information in a Photon Program under U.S. Army Research Office Grant No.\ W911NF-10-1-0404.


\begin{thebibliography}{2}
\bibitem{footnote}Here we shall treat lensless pseudothermal ghost imaging in transmission, because that is the case for which Ref.~4 develops the nonlocal two-photon interference interpretation in the absence of turbulence.  For independent turbulence on all propagation paths and lensless pseudothermal ghost imaging in reflection, Hardy and Shapiro\cite{Hardy} have shown that the turbulence-induced spatial resolution limit is the same as that found for ghost imaging in transmission.
\end{thebibliography}
\end{document}